\documentclass[prb,11pt,tightenlines]{revtex4} 

\usepackage[bookmarks,pdfhighlight=/O,colorlinks=false,pdfstartview=FitH]{hyperref}
\usepackage{graphicx,color}
\usepackage{natbib}
\usepackage{amsmath,amssymb,mathrsfs,slashed,bm,bbm}
\newcommand{\bvec}[1]{\ensuremath{\boldsymbol{#1}}}
\newcommand{\R}{\ensuremath{\mathbb{R}}}
\newcommand{\dd}{\ensuremath{\mathrm{d}}}
 
\begin{document}

\title{Comment on ``Defining the electromagnetic potentials''}

\author{Hendrik van Hees}
\email{hees@itp.uni-frankfurt.de} 
\affiliation{Institut f{\"u}r Theoretische Physik, Goethe-Universit{\"a}t
  Frankfurt, Max-von-Laue-Str.\ 1, D-60438 Frankfurt am Main, Germany}

\date{\today}

\begin{abstract}
  In this comment it is shown that the argument for a unique
  determination of the electromagnetic potentials in classical
  electrodynamics in \cite{Davis:2020_em_potentials} is flawed. To the
  contrary the ``gauge freedom'' of the electromagnetic potentials has
  proven as one of the most important properties in the development of
  modern physics, where local gauge invariance with its extension to
  non-Abelian gauge groups is a key feature in the formulation of the
  Standard Model of elementary particles in terms of a relativistic
  quantum field theory.
\end{abstract}

\maketitle

\section{Introduction}

In \cite{Davis:2020_em_potentials} the author claims that contrary to
the standard treatment of the electromagnetic potentials in all
textbooks like, e.g.,
\cite{somm3,jackson1998classical,Griffiths:2017_intro_cl_edyn} on classical
Maxwell theory the potentials are to be chosen as those of the Coulomb
gauge. As shall be argued in the following, this is not only
mathematically wrong but also misleading from a physical (as well as
didactic) point of view since the gauge invariance of electromagnetism
is the paradigmatic example for a local gauge symmetry demonstrating a
general important concept for the formulation of the Standard Model of
elementary particle physics, describing all hitherto observed elementary
particles and their interactions in terms of a (renormalizable)
relativistic quantum field theory. In this sense the claim of any
fundamental a-priori preference for any specific gauge is also highly
misleading from a pedagogical point of view.

From the theoretical-physics point of view it is quite commonly accepted
for a long time that the fundamental laws governing the realm of
classical electrodynamics are the ``microscopic'' Maxwell equations in
differential form (for the historical context see, e.g., the remark in
the introductory chapter in \cite{somm3}),
\begin{alignat}{2}
\label{1}
&\bvec{\nabla} \times \bvec{E}+\frac{1}{c} \partial_t \bvec{B}=0,\\
\label{2}
& \bvec{\nabla} \cdot \bvec{B}=0,\\
\label{3}
&\bvec{\nabla} \times \bvec{B} -\frac{1}{c} \partial_t
\bvec{E}=\frac{1}{c} \bvec{j} \\
\label{4}
&\bvec{\nabla} \cdot \bvec{E}=\rho,
\end{alignat}
where the Heaviside-Lorentz system of units has been used, which is
more convenient for theoretical purposes than the SI units used in
\cite{Davis:2020_em_potentials}.

\section{Helmholtz's theorem}
\label{sec.2.helmholtz}

First it is important to note that Helmholtz's theorem is applicable to
time-dependent as well as to time-independent vector fields and states
in a quite general form \cite{somm2,Blumenthal:1905_helmholtz_zerlegung}
that if a vector field $\bvec{V}$ and its first derivatives, which are
themselves differentiable, vanish at infinity, it can be decomposed as
$\bvec{V}=\bvec{V}_1+\bvec{V}_2$ such that
$\bvec{\nabla} \times \bvec{V}_1=0$ and
$\bvec{\nabla} \cdot \bvec{V}_2=0$. With given source,
$\bvec{\nabla} \cdot \bvec{V}=\bvec{\nabla} \cdot \bvec{V}_1=J$, and curl
$\bvec{\nabla} \times \bvec{V}=\bvec{\nabla} \times \bvec{V}_2=\bvec{C}$,
the decomposition is unique up to additive constants for the vector
fields $\bvec{V}_1$ and $\bvec{V}_2$. In the following we tacitly assume
the conditions on the fields needed for the following manipulations
being justified.

Further there are theorems that any curl-free vector field can be
written (at least in any simply connected region of space) as the
gradient of a scalar potential, i.e., $\bvec{V}_1=-\bvec{\nabla}\Phi$,
where $\Phi$ is unique up to a constant and any source-free vector field
can be written as the curl of a vector potential
$\bvec{V}_2=\bvec{\nabla} \times \bvec{A}$, and of course $\bvec{A}$ is
unique only up to an arbitrary gradient field, and this freedom can be
used to impose one constraint condition (``gauge condition'') on
$\bvec{A}$.

Defining $\bvec{\nabla} \cdot \bvec{V}=J$ and $\bvec{\nabla} \times
\bvec{V}=\bvec{C}$ we have
\begin{equation}
\label{5}
\bvec{\nabla} \cdot \bvec{V}=\bvec{\nabla} \cdot \bvec{V}_1=-\Delta \Phi=J,
\quad \bvec{\nabla} \times \bvec{V} = \bvec{\nabla} \times
\bvec{V}_2=\bvec{\nabla} \times (\bvec{\nabla} \times \bvec{A}) =
\bvec{\nabla} (\bvec{\nabla} \cdot \bvec{A}) - \Delta \bvec{A}.
\end{equation}
Since we know from electrostatics how to solve the Poisson equation with
the Green's function of the Laplace operator (here for ``free space'',
i.e., without boundary conditions for Cauchy or Neumann problems as
needed in electrostatics at presence of conductors or dielectrics), it
is convenient to impose the additional constraint
$\bvec{\nabla} \cdot \bvec{A}=0$ (``Coulomb gauge condition''), such
that
\begin{equation}
\label{6}
\Phi(\bvec{x})=\int_{\R^3} \dd^3 x' \frac{J(\bvec{x}')}{4 \pi
  |\bvec{x}-\bvec{x}'|}, \quad \bvec{A}(\bvec{x})=\int_{\R^3} \dd^3 x'
\frac{\bvec{C}(\bvec{x}')}{4 \pi |\bvec{x}-\bvec{x}'|}.
\end{equation}
These formulae can be proven using Green's theorem. Then one has
\begin{equation}
\label{7}
\bvec{V}_1=-\bvec{\nabla} \Phi+\bvec{V}_1^{(0)}, \quad
\bvec{V}_2=\bvec{\nabla} \times \bvec{A} + \bvec{V}_2^{(0)},
\end{equation}
where $\bvec{V}_1^{(0)}=\text{const}$ and
$\bvec{V}_2^{(0)}=\text{const}$. Of course, if it is known that
$\bvec{V}_1$ and $\bvec{V}_2$ vanish at infinity, e.g., if $J$ and $\bvec{C}$ have
compact support, these constants are both determined to
vanish given the potentials (\ref{6}).

As we shall see, however, Helmholtz's decomposition theorem is not of
prime importance to introduce the electromagnetic potentials. For this
it is sufficient that a curl-free vector field can be written as the
gradient of a scalar potential and that a source-free field can be
written as the curl of a vector potential. For a given curl-free vector
field its scalar potential is defined up to an additive constant, and for
a given source-free vector field its vector potential is only determined
up to a gradient of an arbitrary scalar field. In fact, as we shall see,
for the solution of Maxwell's equation with given sources $\rho$ and
$\bvec{j}$ the Helmholtz decomposition theorem is of not too much
practical use. One rather needs a Green's function of the D'Alembert
operator $\Box=\Delta-1/c^2 \partial_t^2$, of which in classical
electrodynamics usually the \textbf{retarded propagator} is the relevant
one (for reasons of causality). 

\section{The electromagnetic potentials}
\label{sect.3.em_pots}

To see, why the claim that there is a preferred or even unique choice of
the electromagnetic potentials is flawed, in this Sect.\ we briefly
summarize the standard textbook procedure in introducing the
electromagnetic potentials and arguing why they are only defined only up
to a gauge transformation.

The electromagnetic potentials are introduced using the homogeneous
Maxwell equations (\ref{1}) and (\ref{2}). Though they have profound
physical meaning, from a mathematical point of view they are merely
constraint conditions on the electric and magnetic fields, but
nevertheless necessary to make the solutions of the complete set of the
initial-value problem of Maxwell's equations unique, which describe the
charge and current densities as the sources of the electromagnetic field
and thus provide the dynamical equations of motion.

The homogeneous Maxwell equations (\ref{1}) and (\ref{2}) imply the
\emph{existence} of a vector and a scalar potential $\bvec{A}$ and
$\Phi$ such that
\begin{equation}
\label{8n}
\bvec{E}=-\vec{\nabla} \Phi -\frac{1}{c} \partial_t \bvec{A}, \quad
\bvec{B}(t,\bvec{x})=\bvec{\nabla} \times \bvec{A}(t,\bvec{x}).
\end{equation}
It is also clear that the potentials are \emph{not uniquely defined} by
the electromagnetic field, $(\bvec{E},\bvec{B})$ since a gauge
transformation to new potentials $\Phi'$ and $\bvec{A}'$,
\begin{equation}
\label{9n}
\Phi' = \Phi + \frac{1}{c} \partial_t \chi, \quad
\bvec{A}'=\bvec{A}-\bvec{\nabla} \chi,
\end{equation}
with an arbitrary scalar field $\chi$ leads to the same electromagnetic
field $(\bvec{E},\bvec{B})$. While $\bvec{E}$ and $\bvec{B}$ are
observable fields, operationally defined as providing the Lorentz force
$\bvec{F}=q (\bvec{E}+\bvec{v} \times \bvec{B}/c)$, the potentials are
not directly observable and only defined modulo a gauge transformation
(\ref{9n}). 

Using (\ref{8n}) in the inhomogeneous Maxwell equations (\ref{3}) and
(\ref{4}) yields
\begin{alignat}{2}
\label{14}
&-\Box \bvec{A} + \bvec{\nabla} \left (\bvec{\nabla} \cdot \bvec{A} +
  \frac{1}{c} \partial_t \Phi \right) = \frac{1}{c} \bvec{j}, \\
\label{15}
&-\Delta \Phi - \frac{1}{c} \partial_t \bvec{\nabla} \cdot \bvec{A} = \rho.
\end{alignat}
Here, the d'Alembert operator is used with the sign convention as in
\cite{Davis:2020_em_potentials}, i.e., $\Box=\Delta-1/c^2
\partial_t^2$. It is clear that these two equations alone do not resolve
the ambiguity in the choice of the potentials since these equations are
of course still gauge invariant, because they are formulated originally
in terms of the Maxwell equations (\ref{3}) and (\ref{4}) involving only
the gauge invariant fields $(\bvec{E},\bvec{B})$. Thus (\ref{14}) and
(\ref{15}) do not provide any constraint for the choice of gauge, i.e.,
we can still impose one constraint on the potentials to facilitate the
solution of the equations (\ref{14}) and (\ref{15}).

A glance at (\ref{14}) immediately shows that a promising choice
for a gauge constraint is the \textbf{Lorenz-gauge condition},
\begin{equation}
\label{16}
\bvec{\nabla} \cdot \bvec{A}_{\text{L}}+\frac{1}{c} \partial_t \Phi_{\text{L}}=0.
\end{equation}
The index L indicates the Lorenz-gauge potentials. Then from (\ref{14})
and (\ref{15}) one finds the inhomogeneous wave equations for
the potentials
\begin{equation}
\label{17}
-\Box \Phi_{\text{L}}=\rho, \quad -\Box \bvec{A}_{\text{L}}=\frac{1}{c} \bvec{j},
\end{equation}
i.e., in the Lorenz gauge the equations for the Cartesian components of
the vector potential decouple from each other as well as from the scalar
potential.

Of course, the inhomogeneous wave equation with a given source is also
not uniquely solvable but one has to impose initial as well as boundary
conditions to make its solution unique, because its solutions are only
determined up to a solution of the homogeneous wave equation, and this
can be constrained by imposing initial conditions as well as boundary
conditions. For the here discussed case of the microscopic Maxwell
equations the boundary conditions are usually imposed at spacial
infinity implied by the physical situation. E.g., one usually has
charges and currents only in a compact spatial region and thus looks for
solutions of the wave equations (\ref{16}) and (\ref{17}) describing
waves radiating outwards from these sources. Indeed, as correctly stated
in \cite{Davis:2020_em_potentials}, also from a causality argument it is
justified to choose the \textbf{retarded solution} for the potentials,
\begin{equation}
\label{19}
\Phi_{\text{L}}(t,\bvec{r}) = \int_{\R^3} \dd^3 r'
\frac{\rho(t-|\bvec{r}-\bvec{r}'|/c,\bvec{r}')}{4 \pi |\bvec{r}-\bvec{r}'|},
\quad \bvec{A}_{\text{L}}(t,\bvec{r}) = \int_{\R^3} \dd^3 r'
\frac{\bvec{j}(t-|\bvec{r}-\bvec{r}'|/c,\bvec{r}')}{4 \pi
  c|\bvec{r}-\bvec{r}'|}.
\end{equation}
The initial condition can then be satisfied by adding an appropriate
solution of the homogeneous wave equations, $\Box \Phi_{\text{L}}=0$ and
$\Box \bvec{A}_{\text{L}}=0$. This of course implies that also the
physical em.\ field $(\bvec{E},\bvec{B})$ is given by retarded
solutions and thus fulfill the demand of causal solutions that the
observable electromagnetic field are ``caused'' by the presence of the
charge and current densities as sources, and the field depends at time $t$ only on the
configuration of these sources at the earlier times
$t_{\text{ret}}=t-|\bvec{r}-\bvec{r}'|/c$.

One should note that this satisfies not only the mere causality
condition but even the more strict condition of ``Einstein causality''
in special relativity, i.e., that an event in Minkowski spacetime can
only be causally connected to events in its past light cone, which
implies that no causal signals can travel faster than the speed of light
in vacuum, $c$. As shown above the electromagnetic field propagates with
this ``limiting speed'' of special relativity.

It is clear, though, that imposing this ``causality constraint'' on the
potentials is not a priori necessary, since only the observable em.\
field $(\bvec{E},\bvec{B})$ needs to be ``causally connected''
functionals of the sources $(\rho,\bvec{j})$. Indeed, since the fields
are given by derivatives of the potentials, the ``causal choice''
(\ref{19}), using the retarded Green's function for the $\Box$-operator,
as the solution of the equations (\ref{14}) and (\ref{15}) implies that
also $(\bvec{E},\bvec{B})$ are retarded solutions and thus fulfill the
causality condition.

It is also important to note that (\ref{19}) provide only a solution to
Maxwell's equations if the Lorenz condition (\ref{15}) indeed is
fulfilled. A simple calculation shows that this is the case, if
the continuity equation,
\begin{equation}
\label{20}
\partial_t \rho + \bvec{\nabla} \cdot \bvec{j}=0,
\end{equation}
i.e., the local form of charge conservation is fulfilled. This is anyway
a necessary integrability condition for the Maxwell equations,
and thus independent of the introduction of the potentials and the
choice of their gauge.

One should also note that the Lorenz-gauge constraint (\ref{16}) does
\emph{not} uniquely determine the potentials, since one can change the
potentials by a gauge transformation with a scalar field $\chi$
according to (\ref{9n}), fulfilling the homogeneous wave equation,
\begin{equation}
\label{21}
\Box \chi=0,
\end{equation}
which has non-zero solutions (even such vanishing at spatial infinity
like, e.g., the spherical wave
$\chi(t,\bvec{x})=\chi \sin[k(c t-|\bvec{r}|)/|\bvec{r}|]$), without
violating the Lorenz-gauge condition (\ref{16}) for the new potentials. This
arbitrariness of course is again irrelevant for the just determined
causal physical fields $(\bvec{E},\bvec{B})$, because these do not
depend on the gauge and are given as retarded integrals over the sources
$(\rho,\bvec{j})$.

From this line of arguments it is already clear that the potentials do
not need to be necessarily retarded solutions to fulfill the causal
connection between sources and physical fields, and thus any other
gauge, which may not allow for entirely retarded solutions is as
justified as the Lorenz gauge.

This is of course, even in an extreme sense, illustrated by the other
most commonly used gauge fixing, the \textbf{Coulomb gauge}. It is
motivated by starting from (\ref{15}) and observing that with imposing
the constraint,
\begin{equation}
\label{22}
\bvec{\nabla} \cdot \bvec{A}_{\text{C}}=0,
\end{equation}
one decouples $\bvec{A}_{\text{C}}$ from the equation for the scalar field, which
now obeys a Poisson equation as in electrostatics (but of course in
general with a time-dependent charge density),
\begin{equation}
\label{23}
-\Delta \Phi_{\text{C}}=\rho
\end{equation}
with the solution
\begin{equation}
\label{24}
\Phi_{\text{C}}(t,\bvec{r})=\int_{\R^3} \dd^3 r' \frac{\rho(t,\bvec{r}')}{4 \pi |\bvec{r}-\bvec{r}'|}.
\end{equation}
This is of course still in some sense a ``causal solution'', as claimed
in the paper, because the integrand only depends on the present time $t$
but not on times $>t$, but it obviously seems to violate ``Einstein
causality'', because the scalar potential at time $t$ is determined by
the present charge configuration at this time $t$ but gets
``instantaneous contributions'' from points $\bvec{r}'$ which may be
arbitrarily far from the observational point $\bvec{r}$. As we shall
see, that is not a problem at all since with the appropriate choice of
solutions one finally ends up with the same retarded physical fields
$(\bvec{E},\bvec{B})$ as with the retarded potentials from the
Lorenz-gauge potentials.

This can be seen by using (\ref{23}) and the Coulomb-gauge condition
(\ref{21}) in (\ref{14}), which gets
\begin{equation}
\label{25}
-\Box \bvec{A}_{\text{C}}=\frac{1}{c} \bvec{j}_{\perp}
\end{equation}
with
\begin{equation}
\label{26}
\bvec{j}_{\perp}(t,\bvec{x}) = \bvec{j}(t,\bvec{x}) - \partial_t
\bvec{\nabla} \Phi_{\text{C}}(t,\bvec{x}) = \bvec{j}(t,\bvec{x}) - \bvec{\nabla}
\int_{\R^3} \dd^3 r' \frac{\partial_t \rho(t,\bvec{r}')}{4 \pi |\bvec{r}-\bvec{r}'|}.
\end{equation}
To see that (\ref{25}) is consistent with the Coulomb-gauge condition
(\ref{22}), i.e., with $\bvec{\nabla} \cdot \bvec{j}_{\perp}$ we again
need the continuity equation (\ref{20}) to rewrite (\ref{26}) to
\begin{equation}
\label{27}
\bvec{j}_{\perp}(t,\bvec{x}) = \bvec{j}(t,\bvec{x}) - \bvec{\nabla}
\int_{\R^3} \dd^3 r' \frac{-\bvec{\nabla}' \cdot \bvec{j}(t,\bvec{r}')}{4
  \pi |\bvec{r}-\bvec{r}'|}.
\end{equation}
Now taking the divergence of this equation indeed gives
\begin{equation}
\label{28}
\bvec{\nabla} \cdot \bvec{j}_{\perp}(t,\bvec{x}) = \bvec{\nabla} \cdot
\bvec{j}(t,\bvec{x}) - \Delta \int_{\R^3} \dd^3 r' \frac{-\bvec{\nabla}' \cdot \bvec{j}(t,\bvec{r}')}{4
  \pi |\bvec{r}-\bvec{r}'|}=0.
\end{equation}
To get retarded solutions for the fields, it seems to be appropriate to
solve (\ref{25}) with the retarded propagator, i.e.,
\begin{equation}
\label{29}
\bvec{A}_{\text{C}}(t,\bvec{r}) = \int_{\R^3} \dd^3 r'
\frac{\bvec{j}_{\perp}(t_{\text{ret}},\bvec{r}')}{4 \pi c|\bvec{r}-\bvec{r}'|}.
\end{equation}
In the following we like to show that indeed the Coulomb-gauge
potentials can be written as a gauge transformation of the retarded
Lorenz-gauge potentials, which of course implies that the physical
fields are the same retarded fields as derived using the Lorenz-gauge
potentials.

For this proof it is convenient to introduce the vector field,
\begin{equation}
\label{30}
\bvec{j}_{\parallel}(t,\bvec{x}) =
\bvec{j}(t,\bvec{x})-\bvec{j}_{\perp}(t,\bvec{x}) =\bvec{\nabla}
\partial_t \Phi_{\text{C}}(t,\bvec{x}) = \bvec{\nabla}
\partial_t \int_{\R^3} \dd^3 r' \frac{\rho(t,\bvec{r}')}{4 \pi
  |\bvec{r}-\bvec{r}'|},
\end{equation}
where we have used (\ref{26}). Then we can write (\ref{29}) in the form
\begin{equation}
\label{31}
\bvec{A}_{\text{C}}(t,\bvec{r}) =\bvec{A}_{\text{L}}(t,\bvec{r}) - \int_{\R^3} \dd^3 r'
\int_{\R} \dd t' \frac{\delta(t'-t+|\bvec{r}-\bvec{r}'|/c)}{4 \pi c
  |\bvec{r}-\bvec{r}'|} \bvec{\nabla}' \partial_{t'} \Phi_C(t',\bvec{r}').
\end{equation}
Integration by parts yields
\begin{equation}
\begin{split}
\label{32}
\bvec{A}_{\text{C}}(t,\bvec{r}) &=\bvec{A}_{\text{L}}(t,\bvec{r}) -
\int_{\R^3} \dd^3 r' \int_{\R} \dd t' \Phi_C(t',\bvec{r}')
\bvec{\nabla}' \partial_{t'}
\frac{\delta(t'-t+|\bvec{r}-\bvec{r}'|/c)}{4 \pi c
  |\bvec{r}-\bvec{r}'|}  \\
&= \bvec{A}_{\text{L}}(t,\bvec{r}) - \frac{1}{c} \partial_t
\bvec{\nabla} \underbrace{\int_{\R^3} \dd^3 r' \int_{\R} \dd t'
\Phi_C(t',\bvec{r}') \frac{\delta(t'-t+|\bvec{r}-\bvec{r}'|/c)}{4 \pi
  |\bvec{r}-\bvec{r}'|}}_{\Psi_{\text{CL}}(t,\bvec{r})}.
\end{split}
\end{equation}
With this first we have
\begin{equation}
\label{33}
\bvec{A}_{\text{C}}(t,\bvec{r}) = \bvec{A}_{\text{L}}(t,\bvec{r})
-\bvec{\nabla} \chi_{\text{CL}}
\end{equation}
with the scalar field defining the gauge transformation from the Lorenz-
to the Coulomb-gauge potentials,
\begin{equation}
\label{34}
\chi_{\text{CL}}=\frac{1}{c} \partial_t \Psi_{\text{CL}}(t,\bvec{r}).
\end{equation}
All we have to show to complete our proof of the gauge equivalence of
the Coulomb-gauge and the Lorenz-gauge potentials is that with this
definition we also fulfill
\begin{equation}
\label{35}
\Phi_{\text{C}}(t,\bvec{r}) = \Phi_{\text{L}}(t,\bvec{x})+\frac{1}{c} \partial_t \chi_{\text{CL}}.
\end{equation}
Now
\begin{equation}
\label{36}
\frac{1}{c} \partial_t \chi_{\text{CL}} = \frac{1}{c^2} \partial_t^2
\Psi_{\text{CL}} = (\Delta-\Box) \Psi_{\text{CL}}.
\end{equation}
The first term is immediately calculated from the definition of
$\Psi_{\text{Cl}}$ in (\ref{32}) since the defining integral is just the
retarded solution of the
inhomogeneous wave equation
\begin{equation}
\label{37}
\Box \Psi_{\text{CL}}=-\Phi_{\text{C}}.
\end{equation}
Further we have, again using the definition of $\Psi_{\text{CL}}$ in
(\ref{32}), integrating by parts, using (\ref{23}) and (\ref{19})
\begin{equation}
\begin{split}
\label{38}
\Delta \Psi_{\text{CL}}(t,\bvec{r}) &= \int_{\R^3} \dd^3 r' \int_{\R} \dd t'
\Phi_C(t',\bvec{r}') \Delta \frac{\delta(t'-t+|\bvec{r}-\bvec{r}'|/c)}{4 \pi
  |\bvec{r}-\bvec{r}'|} \\
&=\int_{\R^3} \dd^3 r' \int_{\R} \dd t'
\Phi_C(t',\bvec{r}') \Delta' \frac{\delta(t'-t+|\bvec{r}-\bvec{r}'|/c)}{4 \pi
  |\bvec{r}-\bvec{r}'|} \\
&=\int_{\R^3} \dd^3 r' \int_{\R} \dd t'
 \frac{\delta(t'-t+|\bvec{r}-\bvec{r}'|/c)}{4 \pi
  |\bvec{r}-\bvec{r}'|} \Delta' \Phi_C(t',\bvec{r}') \\
&\stackrel{\text{(\ref{23})}}{=} -\int_{\R^3} \dd^3 r' \int_{\R} \dd t'
 \frac{\delta(t'-t+|\bvec{r}-\bvec{r}'|/c)}{4 \pi
  |\bvec{r}-\bvec{r}'|} \rho(t',\bvec{r}') \\
&\stackrel{\text{(\ref{19})}}{=} -\Phi_{\text{L}}(t,\bvec{r}).
\end{split}
\end{equation}
Using (\ref{37}) and (\ref{38}) in (\ref{36}) indeed leads to
(\ref{35}), i.e., indeed the Coulomb-gauge potentials, with the choice
of a retarded solution (\ref{29}) of (\ref{25}), are just a gauge
transformation of the retarded Lorenz-gauge potentials and thus the
resulting electromagnetic fields are the same retarded solutions as
derived from the Lorenz-gauge potentials, again underlining the fact
that two sets of em.\ potentials connected by a gauge transformation
with an arbitrary gauge field $\chi$ describe the same physical situation.

The Lorenz-gauge potentials are in some respects more convenient to use
since (a) they admit purely retarded solutions which are usually what is
needed in the physical applications and thus these potentials admit a
manifestly ``causal connection'' with the sources and (b) the
Lorenz-gauge condition is manifestly covariant under Lorentz
transformations since it reads $\partial_{\mu} A^{\mu}=0$ in four-vector
notation (where $(x^{\mu})=(ct,\bvec{x})$, and
$\partial_{\mu}=\partial/\partial x^{\mu}$ are contra- and covariant
four-vector components in Minkowski space).

Nevertheless in some respects the Coulomb gauge has also some
advantages. Among them is that it fixes the gauge more stringently than
the Lorenz-gauge condition. Indeed if we ask for special gauge
transformations,
\begin{equation}
\label{39a}
\bvec{A}'=\bvec{A}_{\text{C}}-\bvec{\nabla} \chi, \quad
\Phi'=\Phi+\frac{1}{c} \partial_t \chi
\end{equation}
such that the Coulomb-gauge condition still holds, this leads to
\begin{equation}
\label{40}
\bvec{\nabla} \cdot \bvec{A}'=\bvec{\nabla} \cdot \bvec{A} - \Delta \chi
=-\Delta \chi \stackrel{!}{=}0.
\end{equation}
This implies that, under the constraint that the new gauge potentials
vanish at spatial infinity as the retarded solutions for localized
sources (i.e., sources with compact spacial support) $\chi=0$, i.e., the
Coulomb-gauge condition is more restrictive than the Lorenz-gauge
condition. I.e., in this sense it provides a \textbf{complete gauge
  fixing} and thus is, e.g., most convenient to quantize the
electromagnetic field in the canonical operator formalism.

It is of course clear that these retarded solutions for
$(\bvec{E},\bvec{B})$ can also be directly derived from the Maxwell
equations (\ref{1}-\ref{4}) without first introducing the
electromagnetic potentials, leading to the so-called \textbf{Jefimenko
  equations}, which are, of course, equivalent to the solutions provided
by the retarded solution of the Lorenz-gauge potentials.

\section{The flaw in Davis's argument}

Given the above standard-textbook derivation of gauge invariance of
classical electrodynamics it is clear that Davis's assertion of being
able to \emph{uniquely} define the electrodynamic potentials must be
flawed. This becomes clear from the paper itself since on the one hand
in Sect.\ 4 he ``proves'' that the Coulomb-gauge potentials are
``uniquely'' determined by an apparently more ``rigorous'' approach to
the Helmholtz decomposition theorem for vector fields, while in Sect.\ 5
he derives the retarded Lorenz-gauge potentials.

First of all the method to introduce certain inverse operators for the
Laplace operator, $\Delta$, and the d'Alembert operator $\Box$ is
mathematically correct. It boils down to define these inverse operators
as integral operators with the usual free-space Green's functions, i.e.,
\begin{equation}
\begin{split}
\Delta^{-1} \phi(t,\bvec{x}) &= -\int_{\R^3} \dd^3 x'
\frac{\phi(t,\bvec{x}')}{4 \pi|\bvec{x}-\bvec{x}'|},\\
\Box^{-1} \phi(t,\bvec{x}) &= -\int_{\R^3} \dd^3 x'
\frac{\phi(t-|\bvec{x}-\bvec{x}'|/c,\bvec{x}')}{4 \pi |\bvec{x}-\bvec{x}'|}.
\end{split} 
\end{equation}
It is also correct that the so-defined operator $\Delta^{-1}$ is unique,
given the physical boundary conditions of sufficiently quickly falling
functions, as is summarized in Sect.\ \ref{sec.2.helmholtz} of this
comment. As detailed in Sect.\ \ref{sect.3.em_pots} the operator $\Box^{-1}$ is not
unique but the retarded propagator is chosen on grounds of the given
causality arguments, i.e., by an additional temporal boundary condition.

In this sense the treatment of the potentials by Davis is mathematically
correct, but it is incomplete, which leads to the wrong assertion that
by using these operator methods the electromagnetic potentials are
uniquely determined. In contradistinction to the physically observable
electromagnetic field, $(\bvec{E},\bvec{B})$, the potentials are not
uniquely defined by the Maxwell equations and appropriate
intial/boundary conditions, but the ``ambiguity'', formalized by the
gauge invariance of the physical observables, is not relevant for
classical electromagnetism as a complete theory of electromagnetic
phenomena.

In the operator-formalism language this is seen as follows. Using
(\ref{2}) it follows that
\begin{equation}
\bvec{\nabla} \times (\bvec{\nabla} \times \bvec{B})=-\Delta \bvec{B}
\end{equation}
and thus, indeed uniquely,
\begin{equation}
\bvec{B} = -\Delta^{-1} \bvec{\nabla} \times (\bvec{\nabla}
\times \bvec{B}) =-\bvec{\nabla} \times \Delta^{-1} (\bvec{\nabla} \times
\bvec{B}).
\end{equation}
The claim that from this the choice
\begin{equation}
\label{39}
\bvec{A}_{\text{C}}=-\Delta^{-1} (\bvec{\nabla} \times \bvec{B}) \qquad \text{(incomplete!)}
\end{equation}
were unique is, however, flawed since, according to Helmholtz's
decomposition theorem applied to the vector potential, $\bvec{A}$,
without specifying $\bvec{\nabla} \cdot \bvec{A}$ the general solution
for the vector potential is
\begin{equation}
\label{41}
\bvec{A}=\bvec{A}_{\text{C}}-\bvec{\nabla} \chi, \quad
\bvec{B}=\bvec{\nabla} \times \bvec{A}=\bvec{\nabla} \times \bvec{A}_{\text{C}}
\end{equation}
with an arbitrary and thus \emph{undetermined} scalar field
$\chi(t,\bvec{x})$. This is precisely the unavoidable ``ambiguity'' in
defining the potentials for a given physical situation due to gauge
invariance. Of course (\ref{39}) is the uniquely defined Coulomb-gauge
vector potential, because it fulfills the Coulomb-gauge condition,
\begin{equation}
\label{42}
\bvec{\nabla} \cdot \bvec{A}_{\text{C}}=-\Delta^{-1} \bvec{\nabla} \cdot
(\bvec{\nabla} \times \bvec{B}) = 0.
\end{equation}
Now it is also clear how to make Davis's arguments complete in the
further steps. First using Faraday's Law (\ref{1})
leads to
\begin{equation}
\vec{\nabla} \times \bvec{E} + \frac{1}{c} \partial_t \bvec{B} =
\bvec{\nabla} \times \left (\bvec{E} +\frac{1}{c} \partial_t
  \bvec{A} \right)=0.
\end{equation}
According to Helmholtz's theorem, this leads to the existence of a
scalar potential, $\Phi$,
\begin{equation}
\label{44}
\bvec{E} + \frac{1}{c} \partial_t \bvec{A} =-\bvec{\nabla} \Phi.
\end{equation}
Using (\ref{4}) and (\ref{41}) and taking (\ref{42}) this results in
\begin{equation}
\label{45}
\bvec{\nabla} \cdot \left (\bvec{E} + \frac{1}{c} \partial_t \bvec{A}
\right) = \rho - \frac{1}{c} \partial_t \Delta \chi =  -\Delta \Phi 
\end{equation}
with the unique solution
\begin{equation}
\label{46}
\Phi = -\Delta^{-1} \rho +\frac{1}{c} \partial_t \chi = \Phi_{\text{C}}
+\frac{1}{c} \partial_t \chi.
\end{equation}
The conclusion is that, contrary to Davis's claims, the electromagnetic
potentials are determined to be the Coulomb-gauge potentials \emph{only}
up to a gauge transformation (\ref{9n}).

The same line of arguments also follows when using the operator
formalism of Sect.\ 5 of Davis's paper, employing $\Box^{-1}$, defined
via the retarded propagator as stated above, instead of $\Delta^{-1}$,
leading to the general solution for the potentials given by the retarded
Lorenz-gauge propagator but again \emph{only} modulo a gauge
transformation,
\begin{equation}
\label{47}
\bvec{A}=\bvec{A}_{\text{L}} - \bvec{\nabla} \chi', \quad
\Phi=\bvec{\Phi}_{\text{L}} + \frac{1}{c} \partial_t \chi'
\end{equation}
with another undetermined scalar field, $\chi'$. Of course the retarded
Lorenz-gauge potentials in the operator prescription are given by
\begin{equation}
\label{48}
\bvec{A}_{\text{L}} = -\Box^{-1} \frac{1}{c} \bvec{j}, \quad
\Phi_{\text{L}}=-\Box^{-1} \rho,
\end{equation}
as correctly stated in Davis's paper, but again also using this strategy
for solving the Maxwell equations for given sources $\rho$ and
$\bvec{j}$ only leads to the determination of the potentials \emph{only}
up to a gauge transformation

\section{Conclusion}

In this comment, we have clarified that the electromagnetic potentials
are \emph{not} uniquely determined by the (relativistic) causality
constraint leading to a unique choice of the potentials, neither as the
retarded solutions of the wave equations for the potentials in Lorenz
gauge nor as the solution of the Coulomb-gauge potentials, as falsely
claimed in \cite{Davis:2020_em_potentials}. We have illustrated that the
``ambiguity'' in the choice of the potentials are mathematical facts
summarizing the standard-textbook approach as well as the operator
approach used in this paper. It should also be emphasized that this
``ambiguity'' is nothing else than gauge invariance of classical Maxwell
theory and is thus irrelevant for the observable electromagnetic fields since
the different electromagnetic potentials related to each other by a
gauge transformation represent the same physics. The causality
constraint, including the more stringent Einstein causality imposed by
the relativistic spacetime structure, has to be imposed only on the
physically observable fields and are not a necessary condition for the
unobservable electromagnetic potentials.

This has been demonstrated by the two standard examples given in most
standard textbooks, the Lorenz gauge, which leads to decoupled
inhomogeneous wave equations for the scalar and the components of the
vector potential, and for them thus the (Einstein) causality condition
can be fulfilled by using the retarded propagator of the d'Alembert
operator, leading to retarded potentials and thus also to a retarded
electromagnetic field. The retardation is given by the speed of light as
to be expected from a massless field like the electromagnetic field. Thus
these retarded solutions obey both the causality and the more stringent
Einstein causality constraints as it must be for a classical
relativistic field theory.

The other ``extreme choice'' with regard to retardation is the Coulomb
gauge, which leads to an instantaneous solution for the scalar potential
and in turn to a wave equation for the vector potential with a nonlocal
distribution to the source. As it turns out, both the scalar and the
vector potential contain non-retarded, instantaneous
contributions. However, the vector potential in the Coulomb gauge
consists of the sum of the retarded Lorenz-gauge vector potential and a
gradient field. This immediately implies that the magnetic field
$\bvec{B}$ calculated from the Coulomb-gauge potential is the same
retarded field as calculated from the retarded Lorenz-gauge vector
potential. Then we have demonstrated that the resulting gauge field also
connects the instantaneous Coulomb-gauge scalar potential with the
retarded Lorenz-gauge scalar potential in the way described by the
corresponding gauge transformation such that also the electric field
turns out to be the same retarded field as calculated from the
Lorenz-gauge potentials.

Of course one can also solve the Maxwell equations without introducing
the potentials, leading to inhomogeneous wave equations, for which the
unique physical choice is the retarded solution due to the usual
causality arguments given above. Here the operator formalism is indeed
more convenient than the standard textbook approach in leading to
\emph{unique} solutions for the physical fields, $(\bvec{E},\bvec{B})$,
the so-called \textbf{Jefimenko equations}. This is an elegant
demonstration of the uniqueness of solutions of the initial-value
problem (together with the appropriate spatial boundary conditions) of
the Maxwell equations in terms of the physical fields, which is usually
shown using the energy expression for the electromagnetic field, as,
e.g., in \cite{somm3}.

It is also interesting that one can choose a different class of gauge
constraints, the so-called ``velocity gauges'' such that the em.\
potentials contain retarded contributions which, however, propagate not
with the speed of light but with arbitrary speeds, even larger than the
speed of light. Of course, these fields are not observable either but lead
again to the same physical retarded solutions for the em.\ field as it
must be \cite{Jackson:2002rj}. 

Further it should be kept in mind that the gauge invariance of classical
electrodynamics is the key to use it in a consistent way in both the
semiclassical description of electromagnetic interactions in
non-relativistic quantum mechanics (i.e.,\ with the charged particles
described in terms of quantum mechanics but the electromagnetic fields
kept at the classical level) or in the full relativistic local quantum
field theory, i.e., QED.

Already the analysis of the unitary representations of the proper
orthochronous Poincar{\'e} group for free massless spin-1 fields leads
to the \emph{necessity} of formulating it as a quantized Abelian gauge
theory to avoid the appearance of continuous intrinsic polarization-like
degrees of freedom, which are indeed not observed in Nature
\cite{sexl01,wein1}.

Contrary to some claims in the literature (see, e.g.,
\cite{Reiss:2017_gauge-restrictions,Reiss:2019_light-matter-int})
neither classical electrodynamics nor quantum theory enables a unique
definition of the electromagnetic potentials, but to the contrary the
gauge symmetry is a \emph{necessary} feature for the consistency of the
description of the electromagnetic interaction in the quantum realm.

\acknowledgments 

I thank the anonymous referee of this comment having drawn my attention
to Refs.\
\cite{Reiss:2017_gauge-restrictions,Reiss:2019_light-matter-int}.


\end{document}